\documentclass[twocolumn]{aastex62}
\usepackage{amsmath, amssymb}
\graphicspath{{./}{figures/}}
\newcommand{\hMpc}{$h^{-1}\,$Mpc}
\newcommand{\lya}{Lyman-$\alpha$ \,}
\newcommand{\lyb}{Lyman-$\beta$ \,}

\newcommand{\sk}[1]{}

\shorttitle{Implication of the Shape of the EDGES Signal for the 21 cm Power Spectrum}
\shortauthors{Kaurov et al.}
\begin{document}


\title{Implication of the Shape of the EDGES Signal for the 21 cm Power Spectrum}

\correspondingauthor{Alexander A. Kaurov}
\author{Alexander A. Kaurov}
\author{Tejaswi Venumadhav}
\author{Liang Dai}
\altaffiliation{NASA Einstein Fellow}
\author{Matias Zaldarriaga}
\affiliation{Institute for Advanced Study, 1 Einstein Drive, Princeton, NJ 08540, USA}
\email{kaurov@ias.edu}

\begin{abstract}
We revisit the 21 cm power spectrum from the epoch of cosmic dawn in light of the recent EDGES detection of the 21 cm global signal at frequencies corresponding to $z\sim20$.
The shape of the signal suggests that the spin temperature of neutral hydrogen was coupled to the kinetic temperature of the gas relatively rapidly ($19\lesssim z \lesssim 21$).
We therefore consider models in which the UV photons were dominantly produced in the rarest and most massive halos ($M\gtrsim 10^9M_\odot$), since their abundance grows fast enough at those redshifts to account for this feature of the signal.
We show that these models predict large power spectrum amplitudes during the inhomogeneous coupling, and then inhomogeneous heating by CMB and \lya photons due to the large shot noise associated with the rare sources. 
The power spectrum is enhanced by more than an order of magnitude compared to previous models which did not include the shot noise contribution, making it a promising target for upcoming radio interferometers that aim to detect high-redshift 21 cm fluctuations.
\end{abstract}


\section{Introduction} \label{sec:intro}

At redshifts $z\sim 20$--$30$, before the epoch of cosmic dawn, the intergalactic medium (IGM) was colder than the cosmic microwave background (CMB), and the spin temperature of the 21 cm line of neutral hydrogen was coupled to the CMB temperature. Subsequently, the spin states were driven toward equilibrium with the thermal motion of the gas due to repeated scattering of the UV radiation from the first stars within the Ly$\alpha$ resonance \citep{1952AJ.....57R..31W,1959ApJ...129..536F}. This process lowers the spin temperature and leads to an absorption feature in the global radio background\footnote{See \citet{2012RPPh...75h6901P} for a review of the physics of the cosmic dawn and the 21 cm line.}. 


Recent results from the Experiment to Detect the Global Epoch of reionization Signature (EDGES) suggest that this transition happened at $z \sim 20$ \citep{2018Natur.555...67B}. Surprisingly, the reported absorption profile (see top panel in Figure \ref{fig:main}) is characterized by abrupt edges and a flattened bottom, which are not seen in any of the prior theoretical models~\citep{2017MNRAS.472.1915C}. Although this distinctive shape may have been affected by the choice of the basis functions used to model the foreground, it is sufficiently intriguing to motivate us to explore extreme scenarios of cosmic \lya coupling.

 
Many studies that immediately followed the EDGES detection have focused on the depth of the absorption feature ($0.5^{+0.5}_{-0.2}$ K at $99 \%$ confidence level), which is a factor of $\sim 2.5$ larger than the value in any of the standard models of the Cosmic Dawn. 
Attempts to explain the size of the signal include either new physics~\citep{2018Natur.555...71B,2018arXiv180210094M,2018arXiv180302804B,2018arXiv180303091B,2018arXiv180309739L,2018arXiv180307048P,2018arXiv180303245F,2018arXiv180309697H} or astrophysics~\citep{2018arXiv180207432F,2018arXiv180301815E}. At the present time, it is still unclear whether all these proposed explanations are physically possible and/or consistent with other measurements with many of the proposals already disfavored by more careful analysis.


In this study we ignore the depth of the absorption signal, and focus on its shape -- the sharp boundary and subsequent flat bottom indicate that \lya photons flooded most of the universe within a small fraction of a Hubble time ($\delta z \sim 2$ at redshift $z\sim 20$), followed by an extended period ($\delta z \sim 2$) with little to no heating before the gas temperature quickly rose and diminished the absorption signal. Our main objective is to see what observational consequences this feature of the signal has on the expected 21 cm power spectrum. Whatever mechanism might be responsible for the anomalous depth of the signal could also change other characteristics of it, so one might view our study independently of the EDGES results as an exploration of a part of parameter space not yet considered.


In \S\ref{sec:masscuts} we argue that the EDGES signal implies that sources of UV emission are hosted in very massive and rare halos. Then, in \S\ref{sec:numerical} we describe our model of inhomogeneous coupling and heating (ignoring X-ray heating). In \S\ref{sec:results} we discuss the observational consequences of our model, and in \S\ref{sec:comparison} we briefly compare our results with other studies. Throughout the paper we use the Planck 2015 cosmological parameters \citep{2016A&A...594A..13P}.

\section{Order of magnitude estimates}
\label{sec:masscuts}

\begin{figure}
\includegraphics[width=\columnwidth]{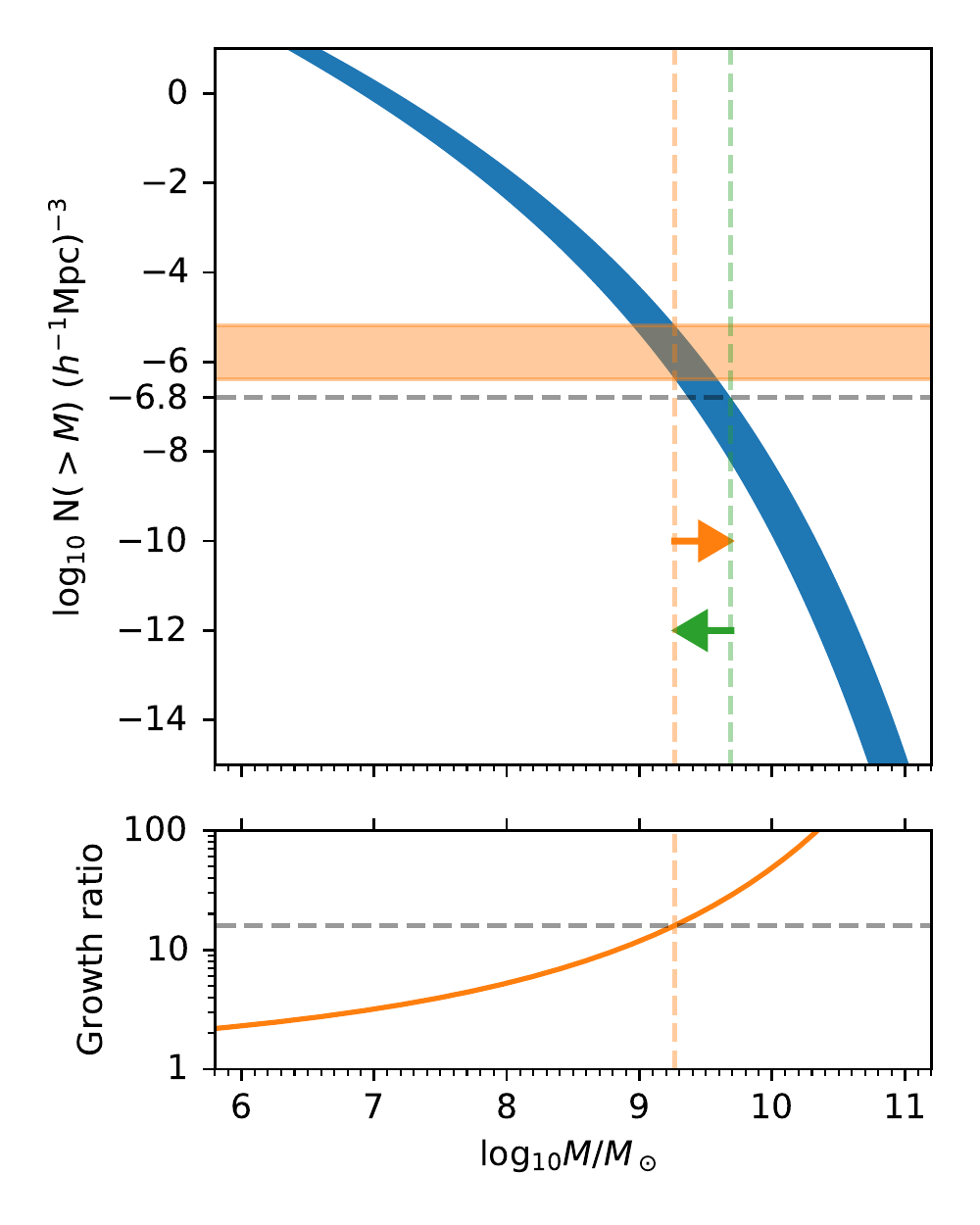}
\caption{\label{fig:growth} Constraints on the lightest masses for the dark matter halos that efficiently produce UV radiation. \textit{Top panel:} Blue filled region corresponds to the evolution of the cumulative mass function between redshifts 19 and 21 assuming the mass function from \citet{2001MNRAS.323....1S}. \textit{Bottom panel:} Growth ratio in the number density of halos with mass exceeding $M$ between $z=19$ and $z=21$. The first cut imposes a minimal number density $N \gtrsim 10^{-6.8}\,(h^{-1}\,{\rm Mpc})^{-3}$ which puts an upper limit on the halo mass $\log_{10}(M/M_\odot) \lesssim 9.7$ (green arrow). The second cut is placed on the growth ratio between the two redshifts to be at least factor of 16 (see bottom panel), which translates to halos more massive than $\log_{10}(M/M_\odot) \gtrsim 9.3$ (orange arrow). See \S\ref{sec:masscuts} for further discussion.}
\end{figure}

\begin{figure*}
\includegraphics[width=\textwidth]{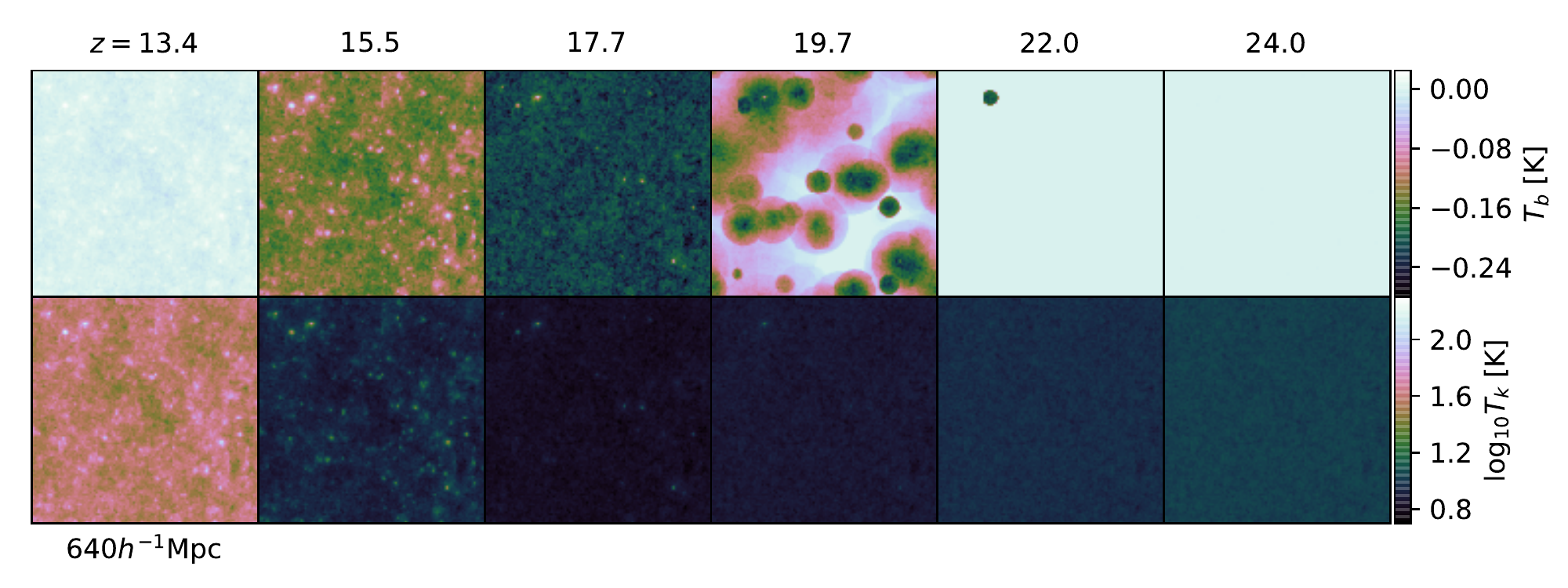}
\caption{\label{fig:grid} The inhomogeneous brightness temperature (top row) and the kinetic gas temperature (bottom row) at six characteristic epochs in (640\hMpc)$^3$ box for our fiducial model with $\log_{10}(M_{\rm  thr}/M_\odot)=9.4$. From right to left: 
at $\bf{z\sim24}$ no sources formed yet, and the temperature follows density perturbations that are very small; 
at $\bf{z\sim22}$ the first coupling bubble starts to form, the kinetic temperature is still almost uniform, and the power spectrum (PS) of 21 cm is dominated by one bubble; 
at $\bf{z\sim19.7}$ we have the intermediate stage of coupling with a lot of bubbles, and the PS is dominated by the shot noise; 
at $\bf{z\sim17.7}$ the universe is coupled everywhere and not heated yet, the PS reaches local minimum; 
at $\bf{z\sim15.5}$ the inhomogeneous heating kicks in and produces, again, signal in the PS dominated by the shot noise; 
finally, at $\bf{z\sim13.4}$ the number of sources is high enough to produce almost uniform heating.}
\end{figure*}

The sharpness of the brightness temperature drop between $z=21$ and $19$ (see Figure \ref{fig:main}) implies that the \lya background grows dramatically during this time. In order to estimate by how much the \lya background needs to increase we perform the following estimate. Given that the ratio of the r.m.s. of the noise to the amplitude of the signal is $0.025/0.5 \sim 5\%$ \citep{2018Natur.555...67B} we can definitively say that the amplitude grew from $20\%$ and $80\%$ of its maximum in $\delta z \lesssim 2$. Since the brightness temperature is proportional to $x_\alpha/(1+x_\alpha)$, where $x_\alpha$ is the \lya coupling coefficient which is proportional to the \lya background, we can conclude that the \lya background grew by factor of $\gtrsim 16$.

Such a rapid growth within a narrow redshift range can be associated only with dramatic changes in certain properties of the UV sources. One possibility is that the physics of star formation abruptly changed, but it is difficult to justify why it occurred within this particular narrow redshift interval. Another plausible explanation, that we adopt, has to do with the abundance of the host halos of the sources. At any given redshift, the abundance of halos at the massive end of the mass function grows exponentially; from $z\sim 21$ to $19$ the halos whose abundance increases by more than a factor of 16 satisfy $\log_{10}(M/M_\odot) \gtrsim 9.3$ (see orange boundary in Figure \ref{fig:growth}). The rapid evolution of the 21cm signal implies that the UV sources reside in these massive halos.

On the other hand, the UV sources have to be sufficiently abundant for a significant fraction of the universe to be coupled. Given that the light travel distance between $z=21$ and $19$ is $\sim 113$ \hMpc, sources should have a number density greater than $10^{-6.8}$ halos per (\hMpc)$^3$. This constraint corresponds to halo masses $\log_{10}(M/M_\odot) \lesssim 9.7$ (see green boundary in Figure \ref{fig:growth}).

In summary, the two mutually compatible constraints provide us with a mass range $9.3 \lesssim \log_{10}(M/M_\odot) \lesssim 9.7$ for the lightest halos that efficiently formed the first stars. Note that these halos are significantly more massive than the widely accepted minimum halo mass of $\sim 10^6 M_\odot$ for the first star formation \citep{2013RPPh...76k2901B}. In the next section, we explore the hypothesis of rare UV sources and study how such a scenario can affect the 21 cm power spectrum.

\section{Inhomogeneous coupling}
\label{sec:numerical}

The dynamic range of the problem (halos of $\sim10^9M_\odot$ separated by 100 \hMpc) does not allow us to run a full N-body dark matter simulation. We instead use an approximate method to populate the volume with halos based on the assumption of log-normal probability density function of galaxy and matter density fields \citep{1991MNRAS.248....1C,2017arXiv171205834H}. 

We adopt a $256^3$ mesh in a (640 \hMpc)$^3$ volume, generate the initial conditions and evolve them with the Zeldovich approximation to the last epoch of our simulation at $z=13$. We then populate the volume with halos using the mass function and the bias prescription from \citet{2001MNRAS.323....1S}. In order to generate halo catalogs at higher redshifts, we modify the halo catalog at $z=13$ by reducing the masses to match the mass function at any given redshift (similar to the abundance matching technique). In other words, we assume gradual growth of all halos through accretion with the rate proportional to their mass. We consider redshifts in range $13< z< 30$ with time step $\Delta z = 0.1$.

We assume that sources are hosted only in the dark matter halos with masses above $M_{\rm thr}$, and their emissivity is proportional to their mass\footnote{This assumption is equivalent to a constant star formation efficiency $f_\star$. Replacing it with a more sophisticated model does not dramatically change our results since the flux is dominated by halos in a very narrow mass range.}. The spectral energy distribution is assumed to be flat between \lya and \lyb in terms of number of photons per frequency bin.

At each redshift step and in each cell of the simulation box, we evaluate the \lya background as the sum of the UV radiation from all sources that redshifts into \lya at given location. Knowing the \lya background in each cell as a function of redshift and the local linearly growing overdensity, we solve the thermal history and calculate the brightness temperature of the 21 cm line, taking into account both \lya and CMB heating \citep{2018arXiv180402406V}.

For the sake of comparison, we also run a ``smooth'' simulation in which we distribute the UV sources in all cells according to the clustering bias of their sources, instead of explicitly creating individual halos. This approach assumes the same mass function and bias, and as a result has the same total number of emitted photons as our fiducial simulation. However, by construction this simulation does not incorporate any shot noise.

We can change the duration and the moment of the coupling by adjusting $M_{\rm thr}$ and the normalization of the flux. Our fiducial model shown in Figures \ref{fig:grid}, \ref{fig:main} and \ref{fig:delta2} uses $\log_{10}(M_{\rm thr}/M_\odot)=9.4$. In Figure \ref{fig:comparison} we show that the model with $\log_{10}(M_{\rm thr}/M_\odot)=9.2$ cannot fit the detailed shape of the global signal -- it either couples fast enough but too early, or starts to couple at the right time but does so too slowly.

One possible caveat is that the choice of the halo mass function can change the quantitative result of this study, since different mass functions can significantly vary at high masses. For example, the value of $M_{\rm thr}$ that leads to an order of magnitude change in the abundance over the required period will be slightly different, and the abundance of these halos would change. However, the results we present should remain qualitatively unchanged. 

\section{Results \& Discussion}
\label{sec:results}

\begin{figure}
\includegraphics[width=\columnwidth]{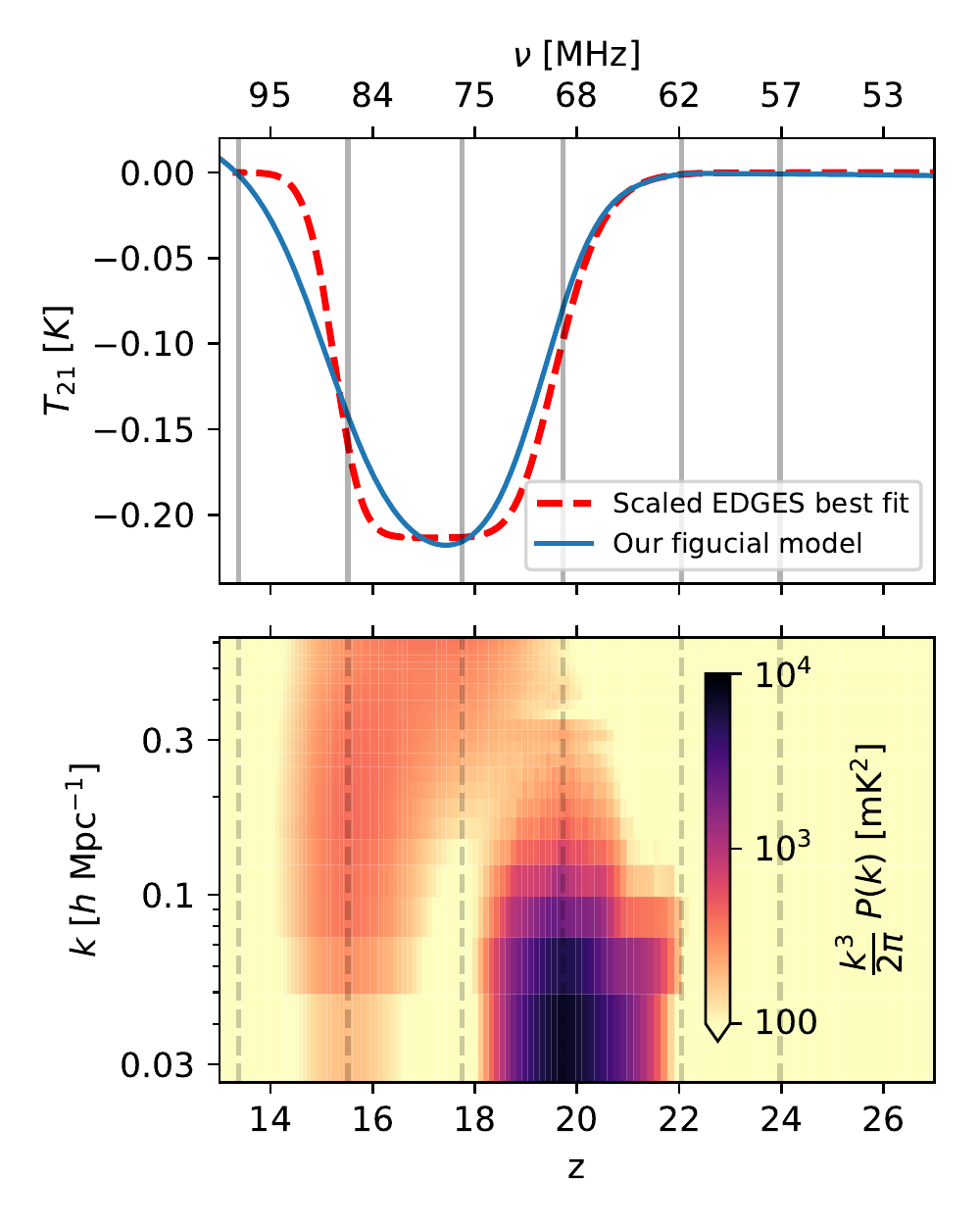}
\caption{\label{fig:main} \textit{Top panel}: the EDGES best fit for 21 cm line brightness temperature scaled by 0.4 is shown with red dashed line, and our fiducial model with $\log_{10}(M_{\rm thr}/M_\odot)=9.4$ is shown with blue solid line. \textit{Bottom panel:} the power spectrum of the 21 cm line at each redshift. Vertical lines correspond to the redshifts that are shown in Figures \ref{fig:grid} and \ref{fig:delta2}.} 
\end{figure}

\begin{figure}
\includegraphics[width=\columnwidth]{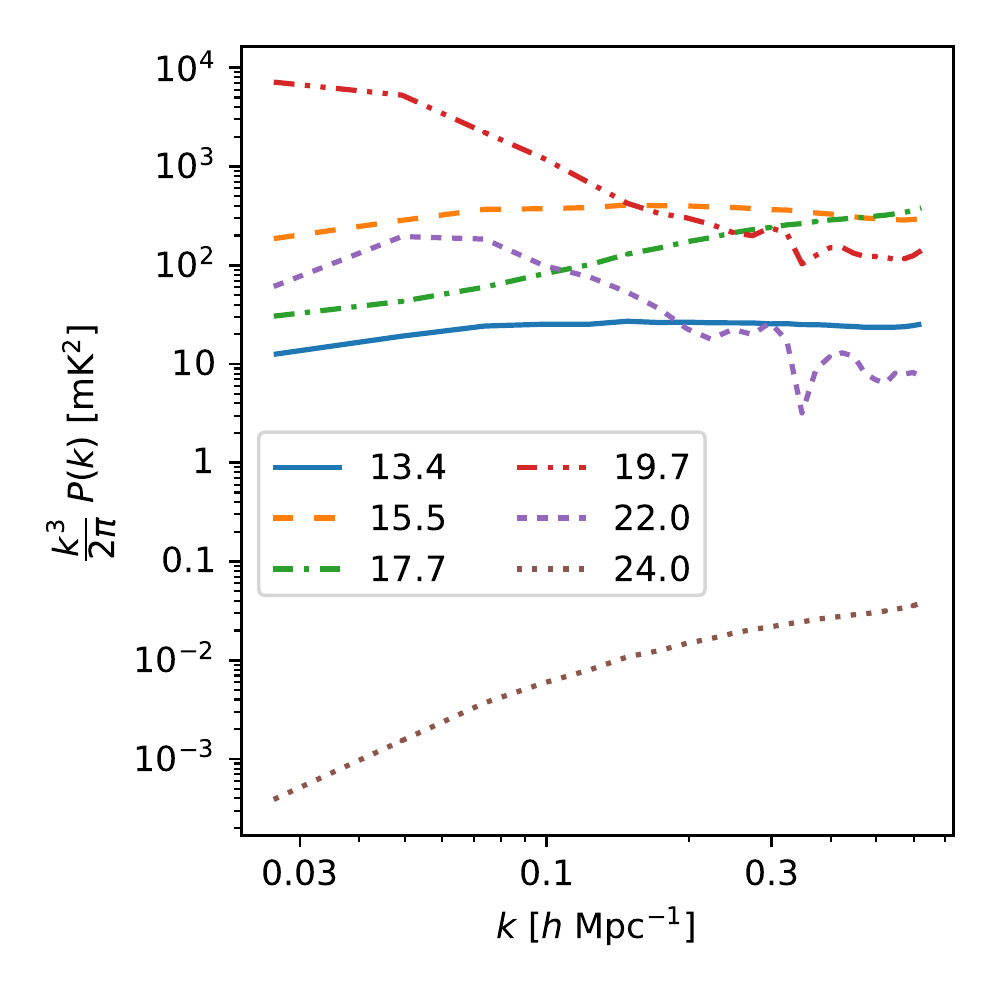}
\caption{\label{fig:delta2} The dimensionless power spectrum at reference redshifts shown in Figures \ref{fig:grid} and \ref{fig:main}. At redshifts 19.7 and 15.5 the power spectrum is dominated by the shot noise.}
\end{figure}

\begin{figure}
\includegraphics[width=\columnwidth]{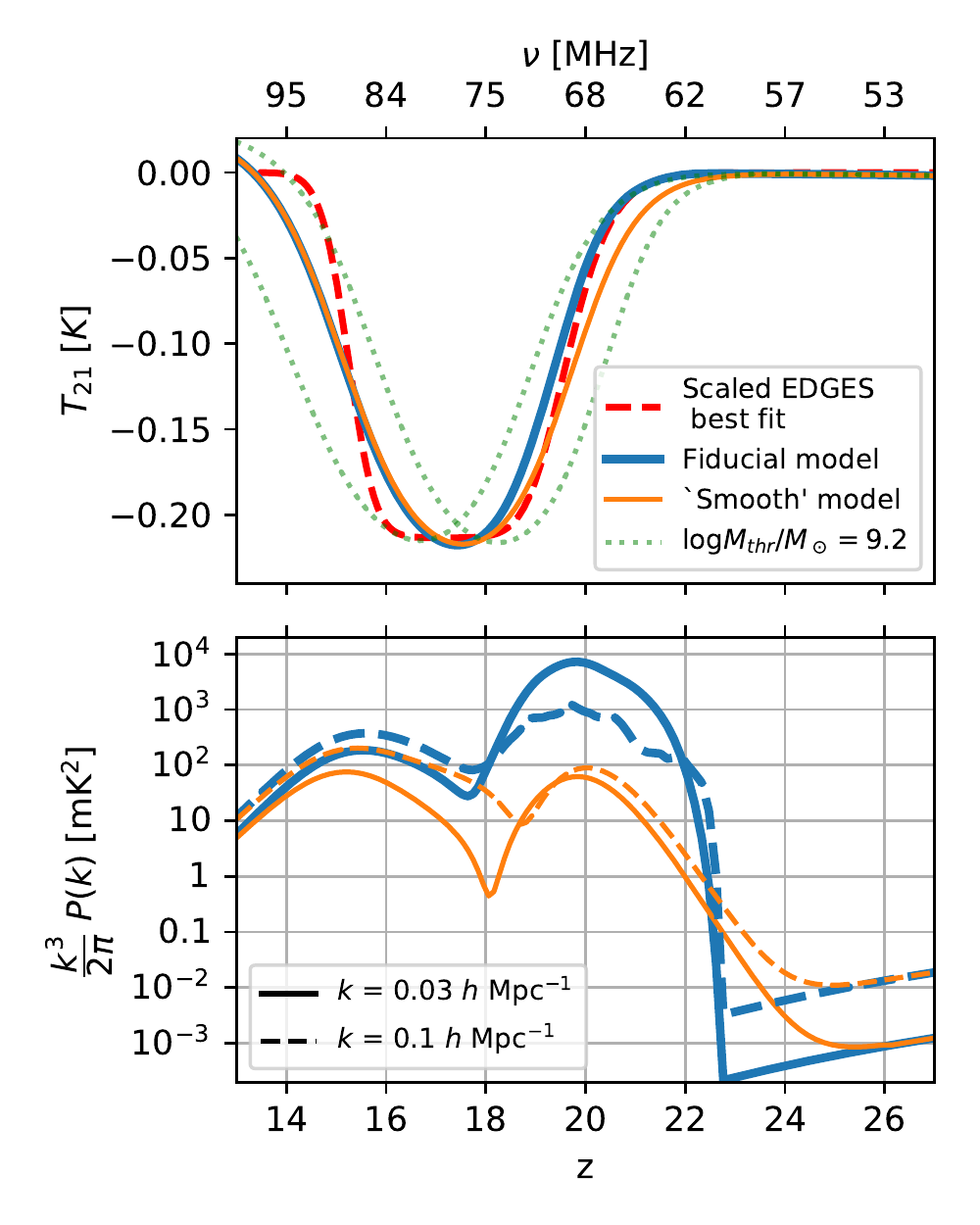}
\caption{\label{fig:comparison} \textit{Top panel}: the EDGES best fit and our fiducial model with $\log_{10}(M_{\rm thr}/M_\odot)=9.4$ repeat Figure \ref{fig:main}. The magenta line shows the `smooth' model with the same number of \lya photons, but without the shot noise. Dotted green lines correspond to the model with $\log_{10}(M_{\rm thr}/M_\odot)=9.2$ and different normalization parameters; they show that they fit into the data much worse comparing to the fiducial model. \textit{Bottom panel:} the fluctuation amplitude of the 21 cm line for $k=0.03$ and $0.1\;h\mathrm{Mpc}^{-1}$ as a function of redshift for the fiducial model and the `smooth' model.} 
\end{figure}

The sharpness of the coupling side of the absorption feature tells us that the sources of the UV photons are likely to reside in the massive halos. For our fiducial model we assume that the majority of star formation happens in dark matter halos with masses $\log_{10}(M_{\rm thr}/M_\odot)>9.4$. In Figure \ref{fig:main} we show this model can produce a global absorption feature similar to the EDGES best fit scaled by a factor of $0.4$, while the models with lower $M_{\rm thr}$ do not produce a sharp feature at a given redshift (see Figure \ref{fig:comparison} for an example).

The bottom panel of Figure \ref{fig:main} shows the evolution of the 21 cm power spectrum, which exhibits multiple local maxima. The dimensionless power spectrum at a number of characteristic epochs is shown in Figure \ref{fig:delta2}. The first peak in the power spectrum at $z\sim 19.7$ corresponds to the shot noise due to the finite number of \lya coupled bubbles (see Figure \ref{fig:grid}). Subsequently, at $z\sim 15.5$ we have the analogous effect due to the CMB heating, whose amplitude is lower since the number of sources by that moment has greatly increased. 

The power spectrum in these peaks is completely dominated by the shot noise. To illustrate this we compare it with that in the ``smooth'' simulation in which photon sources are distributed smoothly across the cells according to the clustering bias, i.e. with no discrete halos. By construction, such a simulation does not include any shot noise. In the bottom panel of Figure \ref{fig:comparison}, we compare the amplitude of the fluctuations in the two models. It can be clearly seen that the shot noise amplifies the peak of \lya coupling by two orders of magnitude, and thus dominates the signal. The model we explored is one with almost the most prominent effect of shot noise; however, shot noise can be important even in much less extreme models. The shot noise during the \lya coupling was studied with a full numerical simulation at much lower redshifts ($z\sim12$) in \citet{2017arXiv170904353K}, in which the effect on the power spectrum was estimated to be at 5-10\%.

The observation of those two peaks in the 21cm power spectrum would be a confirmation of the EDGES results, and shot-noise enhancement of the power spectrum makes it a promising observational target.

Also, in the model we present here, the majority of the stars are located within the few rarest halos, making them exceptionally bright. Therefore, the forthcoming James Webb Space Telescope should be capable of constraining the abundance of these ultra-luminous halos. A recent study by \citet{2018arXiv180303272M} has explored the consequences of the EDGES results on the expected luminosity function of galaxies at high redshifts.

We would like to emphasize that our results on the heating side of the absorption feature may not be as trustworthy for several reasons. First, we have completely neglected X-ray heating, which is believed to be dominant at these redshifts. Second, we have adopted the same prescription for star formation before and after \lya coupling. Star formation could possibly change to some extent in this redshift range due to the formation of the Population II stars and radiative feedbacks. Nevertheless, it is notable that in our setup the \lya and CMB heating alone do not create a flattened bottom in the absorption dip, which would be even more difficult to explain if additional X-ray heating is taken into account.

Throughout this study we have ignored the fact that the observed absorption feature has a larger depth than expected, which requires additional explanations as mentioned in Introduction. Any of the proposed mechanisms to cool the IGM or to ``heat up'' the CMB does not directly contradict our assumptions. On the contrary, they will make the contrast in brightness temperature between the coupled and the non-coupled regions higher, and consequently further amplify the 21cm power spectrum.

\section{Comparison with previous studies}
\label{sec:comparison}

Inhomogeneous \lya coupling has been taken into account by various approaches. In a survey of the parameter space for cosmic dawn by \citet{2017arXiv170902122C}, the authors adopted a grid with $3\,h^{-1}\,\mathrm{Mpc}$ resolution and calculated the total flux generated in each cell assuming a sub-grid model for the local mass function of halos \citep[see the description of the code in][]{2012Natur.487...70V, 2014MNRAS.437L..36F}. This approach is similar to our ``smooth'' simulation and is valid when each cell contains a large number of sources. In the regime implied by the EDGES results, the sources are very sparse, and hence the shot noise necessarily has to be taken into account. For this reason, the amplitude we have found is more than an order of magnitude larger.

The amplitude of the fluctuations we have found for the fiducial model is similar to the one found by \citet{2015MNRAS.447.1806G}, where a dark matter simulation was used to locate halos in a $100\,h^{-1}\,\mathrm{Mpc}$ box with masses resolved down to $\sim 10^9\,M_\odot$ and smaller masses accounted for with a sub-grid model. In contrast to that simulation, we have focused on the rapid coupling process at $z\sim 20$, and to do so we adopted a larger box in order to capture the rarest halos. In result, we were able to explicitly show the shot-noise contribution to the power spectrum.

\software{
\texttt{nbodykit} \citep{2017arXiv171205834H},
\texttt{hmf} \citep{2013A&C.....3...23M}, \texttt{colossus} \citep{2017arXiv171204512D}
}
\begin{acknowledgments}

AK and TV acknowledge support from the Schmidt Fellowship. TV is also supported by the W.M. Keck Foundation Fund. LD is supported at the Institute for Advanced Study by NASA through Einstein Postdoctoral Fellowship grant number PF5-160135 awarded by the Chandra X-ray Center, which is operated by the Smithsonian Astrophysical Observatory for NASA under contract NAS8-03060. M.Z. is supported by NSF
grants AST-1409709 and PHY-1521097, the Canadian Institute for Advanced Research
(CIFAR) program on Gravity and the Extreme Universe and the Simons Foundation Modern Inflationary Cosmology initiative. 

\end{acknowledgments}

\bibliography{refs.bib,references.bib}
\end{document}